\documentclass[11pt]{article}
\usepackage{epsfig,axodraw}
\input epsf.tex
\newdimen\SaveWidth \SaveWidth=\textwidth
\newdimen\SaveHeight \SaveHeight=\textheight
\advance\SaveWidth by -\textwidth
\advance\SaveHeight by -\textheight

\divide\SaveWidth by 2
\divide\SaveHeight by 2
\advance\hoffset by \SaveWidth
\advance\voffset by \SaveHeight

\def\ie{\it i.e.}
\def\eg{\it e.g.}

\def\etc{\it etc.}

\let\badcite=\cite
\def\cite{~\badcite}

\def\slashchar#1{\setbox0=\hbox{$#1$}           
   \dimen0=\wd0                                 
   \setbox1=\hbox{/} \dimen1=\wd1               
   \ifdim\dimen0>\dimen1                        
      \rlap{\hbox to \dimen0{\hfil/\hfil}}      
      #1                                        
   \else                                        
      \rlap{\hbox to \dimen1{\hfil$#1$\hfil}}   
      /                                         
   \fi} 
    \def\slashword#1{\setbox0=\hbox{$#1$}        
  \dimen0=\wd0                                   
   \setbox1=\hbox{/} \dimen1=\wd1                
   \ifdim\dimen0>\dimen1                         
      \rlap{\hbox to \dimen0{\hfil\bf---\hfil}} %
      #1                                         %
   \else                                         
      \rlap{\hbox to \dimen1{\hfil$#1$\hfil}}    
      /                                          
    \fi}                                         %

\catcode`@=11
\newdimen\vbigd@men                             

\def\vbig#1#2{{\vbigd@men=#2\divide\vbigd@men by 2%
   \hbox{$\left#1\vbox to \vbigd@men{}\right.\n@space$}}}
\catcode`@=12

\catcode`@=11
\def\citenum#1{\csname b@#1\endcsname}
\catcode`@=12

\def\dofig#1#2{\centerline{\epsfxsize=#1\epsfbox{#2}}}
\begin{document}
\begin{titlepage}
\rightline{TUHEP-TH-03145}

\bigskip\bigskip

\begin{center}{\Large\bf\boldmath
Can a Nonsymmetric Metric mimic NCQFT 
in $e^+e^- \to \gamma \gamma$ ?\footnotemark \\}
\end{center}
\footnotetext{ This work was supported by the Department of Physics at Tsinghua University and the Chinese Academy of Sciences.
}
\bigskip
\centerline{\bf Nick Kersting$^{a}$,
		 and Yong-Liang Ma$^{b}$}
\centerline{{\it$^{a}$ High Energy Physics Group, Department of Physics}}
\centerline{{ \it Tsinghua University, Beijing P.R.C. 100084 }}
\centerline{{ kest@mail.tsinghua.edu.cn }}
	
\centerline{{\it$^{b}$ Institute of Theoretical Physics }}
\centerline{{\it Chinese Academy of Sciences, Beijing P.R.C. 100080}}
\centerline{ylma@itp.ac.cn}

\bigskip

\begin{abstract}

	In the nonsymmetric gravitational
theory (NGT) the space-time metric $g_{\mu\nu}$ departs from the
flat-space Minkowski form $\eta_{\mu\nu}$ such that it is no longer
symmetric,
$\ie$  $g_{\mu\nu} \ne g_{\nu\mu}$. We find that in the most
conservative such scenario coupled to quantum field theory, which we call 
Minimally Nonsymmetric Quantum Field Theory (MNQFT), there are 
experimentally measurable consequences 
 similar to those from 
noncommutative quantum field theory (NCQFT). This can be expected 
from the Seiberg-Witten map which has recently been interpreted as
equating gauge theories on noncommutative spacetimes with those
in a field dependent gravitational background. 
In particular, in scattering processes such as the pair annihilation
 $e^+e^- \to \gamma\gamma$, 
 both theories make the same striking prediction that
the azimuthal cross section oscillates in $\phi$. However the predicted 
number of oscillations differs in the two theories: MNQFT predicts between
one and four, whereas  NCQFT has no such restriction. 

\bigskip        

\end{abstract}

\newpage
\pagestyle{empty}

\end{titlepage}

\section{Introduction}
\label{sec:intro}

The search for a unification of gravity and quantum field theory over
the last hundred years has led to several promising candidates,
most notably string theories. While these
theories are not at the stage where they can describe physics 
completely at all energies, they can nonetheless make some interesting
predictions at low energies. One such 
prediction\cite{Witten:1986cc,Seiberg:1999vs} is that
the coordinates of space-time $x_\mu$ , when considered as operators
 $\widehat{x}_\mu$, do not
commute:
\begin{equation}
\label{nceqn}
[\widehat{x}_\mu,\widehat{x}_\nu]=i \theta_{\mu \nu}
\end{equation}
Space-time is then described by this theory of noncommutative 
geometry (NCG)\cite{Varilly:1997qg,Girotti:2003at}.
The real antisymmetric tensor $\theta_{\mu \nu}$ parameterizes the
degree of noncommutivity: ordinary commuting space-time is restored in
the
 $\theta_{\mu \nu} \to 0$ limit. When $\theta_{\mu \nu} \ne 0$ 
the theory is Lorentz-violating and subject to severe experimental
constraints on the various components of $\theta_{\mu \nu}$, ranging from Hydrogen spectra, $e^+e^-$ scattering, and various
CP-violating quantities (see\cite{Hinchliffe:2002km} for a review of the
phenomenology). The collection of these constraints implies
that the dimensionful parameters
$\theta_{\mu\nu}$ should not exceed $1~(TeV)^{-2}$\footnotemark; upcoming
particle colliders with center-of-mass energies near or above the TeV
scale will be able to test this bound.\footnotetext{In some considerations in nuclear physics
this limit can be pushed many orders of magnitude stronger,
 however this assumes that
$\theta_{\mu\nu}$ is constant over solar-system 
scales\cite{Mocioiu:2001nz}}
The Lorentz violation in NCG may be viewed as the presence of a
preferred frame of reference in space parameterized by 
$\overrightarrow{\theta}\equiv \epsilon^{ijk}\theta_{jk}$ with
$\epsilon$ being the Levi-Cevita symbol.
 One consequence of this in the noncommutative quantum field
theory (NCQFT) framework
 is that the differential
cross-section of a scattering experiment, suitably binned over
time to take into account the Earth's motion in this preferred frame,
should have an oscillatory
dependence on the azimuthal angle, $\ie$
\begin{equation}
\label{phi}
\frac{d\sigma}{d\phi} \supset A(\cos\phi, \theta_{\mu \nu})
\end{equation}
where $A$ vanishes in the  $\theta_{\mu \nu} \to 0$ limit.
Since the Standard Model prediction for the azimuthal distribution is
flat,  Eqn (\ref{phi}) would be a particularly striking signal of NCG. In Section
\ref{sec:ncg} we review the calculation of one such scattering cross
section, that of $e^+e ^-$ pair annihilation into photons, and demonstrate
the dependence on the azimuthal angle. This dependence arises from
the appearance of terms in the cross section proportional to 
some in- or out-going momenta
contracted into $\theta_{\mu \nu}$, $\ie$ $p^\mu \theta_{\mu \nu} q^\nu$
where $p,q$ are respectively electron and photon momenta, for example.
Such terms depend explicitly on the sine or cosine of the azimuthal angle of the outgoing
photons.   

Since the antisymmetric contraction of momenta $p^\mu \theta_{\mu \nu} q^\nu$ 
in NCQFT is what leads to the
angular dependence in Eqn (\ref{phi}), we may ask whether some other theory
with an antisymmetric object $a_{\mu \nu}$ may also lead to terms
 like $p^\mu a_{\mu \nu} q^\nu$ in the scattering cross section
from which Eqn (\ref{phi}) 
(with $\theta_{\mu \nu}\to a_{\mu \nu}$) follows.
 One candidate which minimally departs from
standard field theory postulates
that the space-time metric  $g_{\mu\nu}$ is not symmetric,
 $\ie$  $g_{\mu\nu} \ne g_{\nu\mu}$. Then the antisymmetric object
 $a_{\mu \nu}$ is  $\frac{1}{2}(g_{\mu\nu}-g_{\nu\mu})$. Such a
nonsymmetric gravity theory (NGT) has appeared in the literature
 previously\cite{Moffat:1995fc}. 
In particular, we may write 
\begin{equation}
\label{g-components}
g_{\mu\nu} = g_{(\mu\nu)} + g_{[\mu\nu]}
\end{equation}
decomposing $g$ into its symmetric and antisymmetric pieces.
The contravariant tensor $g^{\mu\nu}$ is defined as usual:
\begin{equation}
g^{\mu\nu}g_{\mu\rho} = \delta^\nu_\rho 
\end{equation}
As in conventional general relativity with a symmetric metric,
one can define a Lagrangian density
${\cal L} = \sqrt{-g} R$, where $g\equiv det(g_{\mu\nu})$ and
$R$ is the Ricci scalar,  and derive field equations for $g_{(\mu\nu)}$ 
and $g_{[\mu\nu]}$. 

 There has been extensive work
analyzing the effects of $g_{[\mu\nu]}$ for black hole solutions of the
field equations, galaxy dynamics, stellar stability, and other phenomena
of cosmological and astrophysical 
relevance\cite{Moffat:1997cc,Moffat:1995pi,Moffat:1996dq} where
$g_{(\mu\nu)}$ and $g_{[\mu\nu]}$  may be of comparable size.

In the context of particle physics however, we may start with the
assumption that the curvature of
space in the region of interest is small:
\begin{equation}
\label{g-defn}
g_{\mu\nu} \approx \eta_{\mu\nu} +  h_{(\mu\nu)} + a_{[\mu\nu]}
\end{equation}
where $\eta$ is the usual Minkowski metric and the symmetric
and antisymmetric components $h$ and $a$ both\footnotemark \footnotetext{Note that $a_{\mu\nu}$ cannot
be absorbed into $\eta_{\mu\nu}$ or $h$ by a redefinition of coordinates}
satisfy
$a_{\mu\nu}, h_{\mu\nu} \ll 1 ~\forall~ \mu,\nu$.
We further assume that these fields' dynamics are negligable in the
region of interest and we may treat them as background fields.
The effects of the symmetric tensor $h$ on particle physics in this
limit has been studied elsewhere 
(see for example\cite{h-study,Gusev:1998rp,DiPiazza:2003zp}).
 We would like to
focus our attention here on the effects
of the antisymmetric piece  $a_{\mu\nu}$.

In this work we therefore take  $h_{\mu\nu}=0$.
The components of $a_{\mu\nu}$ are undetermined and random under
the sole restriction that  
$a_{\mu\nu} = {\cal O}(\epsilon) \ll 1 ~\forall~ \mu,\nu$. This amounts
to a space-time metric which fluctuates on scales too small for 
experiment to probe. 
Hence $<a_{\mu\nu}>=0$ and ${\cal O}(\epsilon)$ effects do not 
appear in any measurements. However  $<a^2_{\mu\nu}>\ne 0$ and
${\cal O}(\epsilon^2)$ effects
will appear and may have a significant impact. We term this the
 Minimally Nonsymmetric Quantum Field Theory (MNQFT) and will say
 more of it later.

In this paper we demonstrate that both NCQFT and MNQFT predict
 azimuthal differential
scattering cross sections which oscillate in $\phi$.
In Section \ref{sec:ncg} we first present the NCQFT result, then
in Section \ref{sec:ngt} we derive the prediction from MNQFT.  
Section \ref{sec:discuss} discusses the above results, that
their similarity can be expected on some level via the 
Seiberg-Witten map\cite{Seiberg:1999vs}, and
considers whether other experiments may distinguish the two 
theories.

\section{A Short Review of the NCQFT Calculation}
\label{sec:ncg}
As the lowest order contribution to pair annihilation in NCQFT has
already appeared in full detail in the literature (see\cite{Hewett:2000zp})
we only review some essential features of the calculation here.

We first very briefly mention a few fundamental points in the NCQFT theory 
necessary for the calculation. In particular,
the conventional prescription for converting an ordinary quantum field
theory(QFT) into NCQFT is to replace ordinary products between fields with
a certain ``star-product'':
\begin{equation}
\label{star}
(f \star g)(x) \equiv e^{i \theta_{\mu \nu} \partial_{\mu}^{y} \partial_{\nu}^{z}}
		f(y) g(z) \mid_{y=z=x}
\end{equation}
This definition reproduces 
$[x_\mu,x_\nu]_* \equiv x_\mu*x_\nu - x_\nu*x_\mu  = i \theta_{\mu\nu}$
and hence serves to parameterize NCQFT on coordinate space. Other 
features of QFT remain unchanged. In particular we can write the
NCQED action
\begin{eqnarray}
S_{NCQED} &=& \int d^4 x F^{\mu\nu}*F_{\mu\nu} \\
 &=& \int d^4 x F^{\mu\nu}F_{\mu\nu} \nonumber\\
\nonumber \end {eqnarray}
where the second equality follows by integration by parts.
The NCQED field strength is defined by 
$F_{\mu\nu} \equiv \partial_\mu A_\nu - \partial_\nu A_\mu 
 - i[A_\mu ,A_\nu]_* $. Note the cubic and quartic terms in $F$
will introduce 3- and 4-point couplings for the photon. 
One can derive that the star-products in the
 NCQED Lagrangian give new
Feynman rules very similar to those of QED modulo factors of 
$\theta_{\mu\nu}$ contracted into external leg momenta. Computing
the cross-section for pair annihilation in NCQED is therefore straightforward
but more difficult than in QED since there are three distinct diagrams as 
shown in Figure 1.

{
\unitlength=1.3 pt
\SetScale{1.25}
\SetWidth{0.5}      
\scriptsize    
\begin{picture}(90,90)(0,0)
\ArrowLine(0,20)(30,40)
\ArrowLine(30,40)(60,40)
\Photon(30,40)(0,70){4}{4}
\Photon(60,40)(90,70){4}{4}
\ArrowLine(60,40)(90,20)
\Text(45,8)[c]{\large $(a)$}
\end{picture} \
{} \qquad\allowbreak
\begin{picture}(90,90)(0,0)
\ArrowLine(0,20)(30,40)
\ArrowLine(30,40)(60,40)
\Photon(30,40)(80,70){4}{5}
\Photon(60,40)(10,70){4}{5}
\ArrowLine(60,40)(90,20)
\Text(45,8)[c]{\large $(b)$}
\end{picture} \
{} \qquad\allowbreak
\begin{picture}(90,90)(0,0)
\ArrowLine(0,20)(45,30)
\Photon(45,30)(45,50){4}{3}
\Photon(45,50)(90,70){4}{4}
\Photon(45,50)(0,70){4}{4}
\ArrowLine(45,30)(90,20)
\Text(45,8)[c]{\large $(c)$}
\end{picture} \
{} \qquad\allowbreak
}
\begin{center}
Figure 1.
{\it NCQED Feynman diagrams for  $e^+ e^- \to  \gamma \gamma$.}
\end{center}
 From these the authors of\cite{Hewett:2000zp}  found
\begin{eqnarray}
\label{ncg-result}
\frac{d\sigma}{dz d\phi}&=&\frac{\alpha^2}{4s}
\left[\frac{u}{t} + \frac{t}{u} 
- 4 \frac{t^2+u^2}{s^2}\sin^2(\frac{1}{2}k_{1\mu} \theta^{\mu\nu} k_{2\nu})
\right]
 \\
&=& SM -\alpha^2 \frac{t^2+u^2}{s^3}
\sin^2(\frac{s}{2}(\theta^{01}z+\theta^{02}(1-z^2)\cos\phi
   + \theta^{03}(1-z^2)\sin\phi)) \nonumber \\
\nonumber \end{eqnarray}
where ``SM'' is the Standard Model result, $s,t,u$ are the usual Mandlestam variables and $z$ the cosine of the
polar angle in the laboratory center of mass frame. Here the oscillatory dependence on $\phi$ is clear. Note that the number
of full oscillations in $\frac{d\sigma}{dz d\phi}$ as  $\phi$ goes
from $0$ to $2\pi$ does not have a strict upper bound: the 
higher the product of $s$ and $\theta^{0i}$, the more oscillations.

\section{The MNQFT Calculation}
\label{sec:ngt}

We now put the NCQFT result to the side and turn to a completely
different theory, MNQFT. In this section we will see that MNQFT also
leads to an oscillatory cross section.
The starting point of our calculation is the substitution 
$\eta^{\mu\nu} \to g^{\mu\nu} =\eta^{\mu\nu} + a^{\mu\nu}$ in the Lagrangian
for QED:\footnotemark.
 \footnotetext{In the vierbien formalism, we would
take $g_{\mu\nu}=V^\alpha_\mu(x)V^b\eta_\nu(x)\eta^{\alpha\beta}$,
where the vierbiens $V$ relate the general coordinates to some 
normal coordinates erected at $x$ in terms of which the metric
becomes Minkowski. However in the present case this is not possible
as $g_{\mu\nu}$ is not symmetric.}
\begin{equation}
\label{qedlag}
{\cal L} = \sqrt{-g} \left[\overline{\psi}(i\partial_\mu \gamma^\mu-m)\psi
- \frac{1}{4}F^{\mu\nu}F_{\mu\nu} - e\overline{\psi}\gamma^\mu \psi A_\mu
+ \xi R \right]
\end{equation}
where all space-time index contractions are performed with the full
metric $g^{\mu\nu}$, and we hereafter neglect the curvature term $\xi R$.
This is what we take as the minimal prescription for incorporating NGT 
effects into a QFT calculation: just replace the flat-space
metric $\eta^{\mu\nu}$ with the full metric $g^{\mu\nu}$. 
Other terms could enter the Lagrangian in Eqn (\ref{qedlag}) which explicitly depend on
$a^{\mu\nu}$, such as $a_{\mu\nu}F^{\mu\nu}$, and
may of course be generated by quantum effects, but as such they
will be suppressed by loop factors and we hereafter neglect them as
they will not change the qualitative features of our calculation.

 The Feynman propagators for the electron and photon satisfy,
respectively,
\begin{equation}
\begin{array}{l}
\left[i \partial_\mu\gamma^\mu -m \right]S_F(x,x')
= \left[-g  \right]^{-1/2}\delta^n (x-x') \\
\\
\left[g_{\mu\nu}\bigtriangledown^2 \right]D^{\rho\nu}_F(x,x')
= \left[-g  \right]^{-1/2}\delta^\nu_\mu \delta^n (x-x')
\end{array}
\end{equation}
as in general curved spaces. Written in momentum space,
\begin{eqnarray}
S_F(x,x')
= \left[-g  \right]^{-1/2}\delta^n (x-x')
  \frac{p_\mu \gamma^\mu+m}{p^2 - m^2}
 \nonumber\\
 \\
D^{\rho\nu}_F(x,x')
= \left[-g  \right]^{-1/2}\delta^n (x-x')\frac{g^{\rho\nu}}{p^2}
\nonumber \end{eqnarray}
The Dirac equation in curved space is 
$(i\gamma^\mu \partial_\mu -m)\psi=0$, where in our case the gamma
matrices are of the usual 4-dimensional form satisfying
 $\{\gamma^\mu,\gamma^\nu\}=2\eta^{\mu\nu}$ (see Appendix).

As in ordinary QED we have two diagrams which contribute to pair annihilation
 (see Figure 2). These have
combined amplitude 
\begin{eqnarray}
\label{amp}
iM&=&-ie^2\epsilon^*_\mu(k_2)\epsilon_\nu(k_1)\bar{u}(p_2)[\frac{\gamma^\mu(p_1\hspace{-0.2cm}\slash-k_1\hspace{-0.2cm}\slash+m)\gamma^\nu}{(p_1-k_1)^2-m^2}
+\frac{\gamma^\nu(p_1\hspace{-0.2cm}\slash-k_2\hspace{-0.2cm}\slash+m)\gamma^\mu}{(p_1-k_2)^2-m^2}]u(p_1)\\
&=&-ie^2\epsilon_\mu^*(k_2)\epsilon_\nu(k_1)\bar{u}(p_2)[\frac{\gamma^\mu
-k_1\hspace{-0.2cm}\slash\gamma^\nu+2\gamma^\mu\eta^{\nu\alpha}p_{1\alpha}}{-2p_1\cdot
k_1}+\frac{-\gamma^\nu
k_2\hspace{-0.2cm}\slash\gamma^\mu+2\gamma^\nu\eta^{\mu\alpha}p_{1\alpha}}{-2p_1\cdot
k_2}]u(p_1)
\nonumber \end{eqnarray}

{
\unitlength=1.3 pt
\SetScale{1.25}
\SetWidth{0.5}      
\scriptsize    
\begin{picture}(90,90)(-50,0)
\ArrowLine(0,20)(30,40)
\ArrowLine(30,40)(60,40)
\Photon(30,40)(0,70){4}{4}
\Photon(60,40)(90,70){4}{4}
\ArrowLine(60,40)(90,20)
\Text(6,15)[l]{\large $p_1$}
\Text(17,60)[l]{\large $k_1$}
\Text(80,15)[r]{\large $p_2$}
\Text(66,60)[r]{\large $k_2$}
\Text(45,8)[c]{\large $(a)$}
\end{picture} \
{} \qquad\allowbreak
\begin{picture}(90,90)(-70,0)
\ArrowLine(0,20)(30,40)
\ArrowLine(30,40)(60,40)
\Photon(30,40)(80,70){4}{5}
\Photon(60,40)(10,70){4}{5}
\ArrowLine(60,40)(90,20)
\Text(6,15)[l]{\large $p_1$}
\Text(0,60)[l]{\large $k_1$}
\Text(80,15)[r]{\large $p_2$}
\Text(87,60)[r]{\large $k_2$}
\Text(45,8)[c]{\large $(b)$}

\end{picture} \
{} \qquad\allowbreak

}
\begin{center}
Figure 2.
{\it Definitions of momenta in the MNQFT calculation.}
\end{center}

Special care is required in dealing with photon polarization. In general
curved spaces the concept of photon polarization loses meaning,
but in our case the metric is only perturbed slightly from the
diagonal Minkowski form, so we assume we may retain
the implicit definition of polarization in setting $k_\mu \epsilon^\mu = 0$.
We can rewrite Eqn (\ref{amp}) as 
\begin{equation}
iM=\epsilon^*_\mu(k_2)\epsilon_\nu(k_1)M^{\mu\nu} 
\end{equation}
where $M^{\mu\nu}$ contains only momenta variables, Dirac matrices, and
their contractions with $\eta_{\mu\nu}$. The square of this amplitude
summed over photon polarizations and averaged over electron spins is
\begin{eqnarray}
\frac{1}{4}\sum_{spins}|M|^2&=&\frac{1}{4}\sum^2_{i,j=1}
|\epsilon^{i*}_\mu(k_2)\epsilon^j_\nu(k_1)M^{\mu\nu}|^2 \\
&=& \frac{1}{4}\sum^2_{i,j=1}
  \epsilon^{i*}_\mu(k_2)\epsilon^{i}_\rho(k_2)
   \epsilon^{j*}_\sigma(k_1)\epsilon^j_\nu(k_1)
    M^{\mu\nu}M^{\rho\sigma}
\nonumber \end{eqnarray}
Now this squared amplitude has two parts: $M^{\mu\nu}M^{\rho\sigma}$,
which depends only on the 
external momenta, and the polarization product $\epsilon^{i*}_\mu(k_2)\epsilon^{i}_\rho(k_2)
   \epsilon^{j*}_\sigma(k_1)\epsilon^j_\nu(k_1)$, which implicitly
contains factors of the metric $g_{\mu\nu}$ (and hence also  $a_{\mu\nu}$).
In the final calculation only squares (or fourth powers, which we may
neglect in the first approximation) of the elements of
 $a_{\mu\nu}$ such as $a^2_{01}$, $a^2_{13}$, $\etc$
  can appear since any odd power
of some element of $a_{\mu\nu}$ averages to zero by construction. 
Following this prescription, and taking 
$<a^2_{\mu\nu}>= {\cal O}(\epsilon^2)$ for simplicity,
we obtain a spin-averaged squared matrix element of(see Appendix for details)

\begin{eqnarray}
\label{ngt-result}
&&{1\over4}\sum_s|{\cal
M}|^2\nonumber\\
&&=\mbox{SM}\nonumber\\
&&+8\epsilon^2{{\alpha^2}\over{4s}\sin^3\theta}\bigg\{-{1\over4}(1+\cos\theta)^2(1+8\cos2\theta)(\sin\varphi+\cos\varphi)\nonumber\\
&&-\sin\theta\{\cos\theta\sin^2\varphi[\cos\theta(\sin\varphi+\cos\varphi)-\sin^2\theta]\}\nonumber\\
&&+(1-\cos\theta)^2\{2(1-\cos\varphi)\cos^2\varphi(\sin\varphi+\cos\varphi)\}\bigg\}
\end{eqnarray}
As in NCQFT, we see the appearance of terms that depend on the
sine or cosine of the azimuthal angle.
In Figure 3 we plot the resulting differential cross section
against $\phi$ (having integrated over the polar
angle for $0.1<\cos\theta<0.9$). Note that in this particular case
where all the $<a^2_{\mu\nu}>$ are of comparable size the differential
cross section undergoes one full oscillation in $\phi$. This is because 
upon numerically integrating over $\theta$ the $(\sin\phi + \cos\phi)$ term
in Eqn (\ref{ngt-result}) dominates. 
One could adjust the
$<a^2_{\mu\nu}>$ to allow terms with different $\phi$-dependance
to dominate, but since all terms are proportional to either 
$\sin^i\phi$ or $\cos^i\phi$ (i=1..4) only one to four oscillations
are possible. We further observe from Figure 3 that the MNQFT oscillates
{\it about} the SM result.
This contrasts
from the prediction in NCQFT (see Eqn (\ref{ncg-result})) where
the contribution to
$d\sigma/d\phi$ is strictly negative and 
may undergo any number of oscillations.
\begin{figure}[t]
\dofig{3.50in}{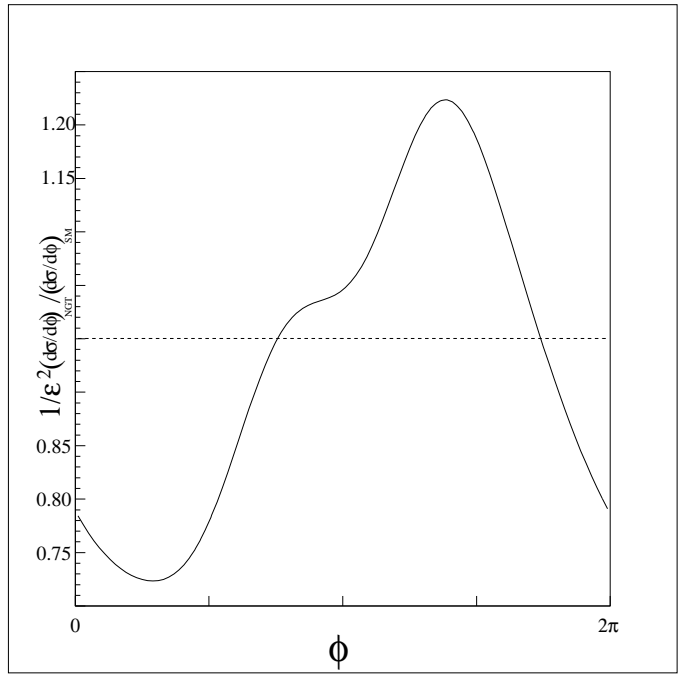}
\caption{ \it Figure 3. 
Ratio of the differential cross section in $\phi$ in MNQFT
to that of SM, for the case where all the elements of $a_{\mu\nu}$ are
of equal average magnitude. Here we have integrated over the polar
angle for $0.1<\cos\theta<0.9$ 
\label{ngt-plot} }
\end{figure}

\section{Discussion}
\label{sec:discuss}
We have seen in the foregoing that both the NCQFT and MNQFT theories
make some similar predictions in high energy processes; we would like to 
remark here that perhaps this is
not so coincidental. The reason why we believe these theories to be
more closely related than at first inspection derives from a well-known
correspondance between ordinary gauge theories on noncommutative spaces and
more complicated gauge theories on ordinary spaces. This is formally known as
the Seiberg-Witten Map\cite{Seiberg:1999vs}. Seiberg and Witten (SW), starting
from the action of the string worldsheet $\Sigma$ in the presence of a
constant ``magnetic field'' $B$, 
\begin{equation}
\label{sw-action}
S= {1\over 4\pi\alpha'}\int_\Sigma \left( g_{ij} \partial_a x^i
\partial^a x^j -2\pi i\alpha' B_{ij}\epsilon^{ab} \partial_a
x^i\partial_b x^j\right)
\end{equation}
restricted to the case where $\Sigma$ is a disc ({\ie} describing
open strings dynamics) obtain two interesting results upon applying
 the boundary conditions on Eqn (\ref{sw-action}): (1)the open strings
feel an ``effective metric'' given by 
$G_{ij} =g_{ij}-(2\pi\alpha')^2 \big(B g^{-1} B\big)_{ij}$;
(2) space-time coordinates do not commute, in that
$ [ x^\mu (\tau),  x^\nu (\tau) ] = i \theta^{\mu\nu}$ where
$\theta^{ij}= 2\pi \alpha' \left({1 \over g+ 2\pi\alpha'
B}\right)^{[ij]}$. Thus we already see that noncommuting coordinates
are related to the space-time metric. Now taking an approximation of 
Eqn (\ref{sw-action}) on a D-brane
where fields are taken to be slowly varying yields the Dirac-Born-Infield (DBI) action\cite{Tseytlin:1999dj},
whose specific form depends on one's regularization scheme: SW showed 
that using
a Pauli-Villars scheme(preserving the gauge symmetry of the
open string gauge fields) leads to a commutative DBI action; however in a
point-splitting regularization scheme one obtains a noncommutative
 DBI action which becomes noncommutative electromagnetism in the
$\alpha' \to 0$ limit.
 Since physics doesn't depend on one's choice of regularization scheme,
Seiberg and Witten proved that these two actions are equivalent
in the sense that there exists a map(via field redefinitions),
the Seiberg-Witten Map, between them.
Recent work has explicitly demonstrated 
this\cite{Garousi:2000,Rivelles:2002ez,Yang:2004vd}:\footnotemark 
\footnotetext{We simplify many details in the ensuing discussion;
the interested reader can pursue the references cited for a
complete treatment}
in a point-splitting scheme in four dimensions one obtains 
\begin{equation}
 \widehat{S}\sim\int d^4 x \widehat{F}_{\mu\nu}
\star   \widehat{F}^{\mu\nu}
\end{equation}
where $\widehat{F}_{\mu\nu} = \partial_\mu \widehat{A}_\nu -  
       \partial_\nu \widehat{A}_\mu
- i \widehat{A}_\mu \star \widehat{A}_\nu + i \widehat{A}_\nu
\star \widehat{A}_\mu$ is the noncommutative field strength; {\ie} this
regularization scheme gives a theory described by NCQFT.
Applying the SW map to the above gives the action
\begin{equation}
 S\sim \int d^4 x\sqrt{\det{(1+ F\theta)}}
\Bigl(\frac{1}{1 + F\theta} F \frac{1}{1 + F\theta} F \Bigr)
\end{equation}
{\ie} an ordinary gauge field theory on a space defined by a 
nonsymmetric metric
\begin{equation}
\label{back-g}
    {\rm g}_{\mu\nu} = \eta_{\mu\nu} + (F\theta)_{\mu\nu}
\end{equation}
We therefore see that the equivalence of gauge theory on a noncommutative 
space and the ordinary theory with a field-dependent background metric 
is a necessary consequence of the SW map.
It is remarkable that our simplistic treatment in the present paper,
 using a minimal coupling ansatz and the metric 
$g^{\mu\nu} =\eta^{\mu\nu} + a^{\mu\nu}$ rather 
than that in Eqn (\ref{back-g}), 
confirms the similarity of the two theories at
 the phenomenological level; that
the exact predictions of NCQFT and MNQFT we have presented for
$e^+e^-$-scattering differ
somewhat may be due to taking the components of $a_{\mu\nu}$ to be
random space-dependent functions,
whereas in 
NCQFT the specific components
of $\theta_{\mu\nu}$ are taken to be fixed and measurable.
We could perhaps therefore view MNQFT as a certain limit of NCQFT where
$\theta_{\mu\nu}$ is no longer a simple constant tensor, but a more
detailed investigation of
this correpsondance will have to wait for a future publication.

From the  analysis of the preceding sections we may conclude that in the pair
annihilation process the predicted number of oscillations in the
azimuthal differential cross section depends on whether space-time is
described by NCQFT or MNQFT. If the former, the number of oscillations 
is unrestricted, whereas the latter predicts between one and four. 
In particular, if less than one oscillation is observed,
MNQFT cannot be responsible and NCQFT would be a candidate explanation with
$s \theta^{0i} < 1$. Moreover, in contrast to MNQFT the 
NCQFT cross section
 is strictly below the SM prediction.
We note further that in NCQFT the number of oscillations 
grows with center of mass
energy as well and in principle one could test this by running a high center 
of mass
$e^+e^-$ linear collider at varying energies if statistics allow
for it.

We believe the foregoing comments will apply to any scattering process,
$\eg$ Moller scattering, Bhahbha scattering, $\etc$
 (see\cite{Kersting:2003ea}) though the NCQFT predictions
will be more robust in processes which do not involve QCD, as the
noncommutative version of QCD has not been as thoroughly developed as
NCQED (however, see\cite{Calmet:2002kf} for encouraging work in this
direction).

Finally, we remark on other types of experiments 
besides those involving high energy scattering. One might expect that low
energy experiments would constrain MNQFT as severely as NCQFT. But due to
the antisymmetry of the metric in MNQFT the definition of distance
$ds^2 = g_{\mu\nu}dx^\mu dx^\nu$ is unchanged and independent of 
 $a_{\mu\nu}$ so that it is not trivial to constrain the theory this
way. Nonrelativistic
quantum mechanics equipped with a Hamiltonian $H=p^2/2m + V(r)$ is
therefore independent of $a_{\mu\nu}$ in contrast to the case in NCQFT
where $\theta_{\mu\nu}$ may have observable effects in the Hydrogen spectrum.
One must go to QED corrections in atomic physics to see the effect of
$a_{\mu\nu}$ but here we expect the effect to be small; the correction
to the anomalous magnetic moment of the muon in MNQFT, for example, is
zero at the one-loop level\cite{Kersting:ngt-g-2}. 
Moreover MNQFT is CP-conserving,
unlike NCQFT which is most strongly constrained by non-observation of
a CP-violating electron electric dipole moment. But in all of the above
experiments the signal of NCQFT or MNQFT will only be a small shift
in a measured quantity such as an energy-level splitting, not as
conspicuous a signal as an oscillating azimuthal cross section, which
we claim to be a superior signal of one theory or the other.
In the realm of cosmology and astrophysics there are many interesting
predictions from NCG and NGT; the former predicts novel features of
the cosmic microwave background spectrum, for example, while the
latter predicts a variety of effects, $\eg$ with respect to black hole solutions of the
Einstein field equations, galaxy dynamics, stellar stability, 
$\etc$
\cite{Moffat:1997cc,Moffat:1995pi,Moffat:1996dq}. Experiments 
in this direction may more strongly distinguish NCG from NGT as the
latter is a purely gravitational effect.

\section*{Acknowledgements}
This work was supported by the
department of Physics at Tsinghua University, the key projects of
Chinese Academy of Sciences and the National Science Foundation of
China (NSFC)

\section*{Appendix}

\subsection*{Gamma Matrices in Our Nonsymmetric Space}
In the most general curved space the Dirac matrices depart from
the usual 4-dimensional form, but in our case, where the metric
differs from Minkowski space by an antisymmetric piece, this is not the case: 
 acting on the Dirc equation on the left with
$(-i\gamma^\nu \partial_\nu -m)$ gives
\begin{eqnarray}
(-i\gamma^\nu \partial_\nu -m)(i\gamma^\mu \partial_\mu -m)\psi=0 
  \nonumber\\
= (\gamma^\nu\gamma^\mu\partial_\nu\partial_\mu+m^2)\psi = 0
  \nonumber\\
=(\frac{1}{2}\{\gamma^\mu,\gamma^\nu\}\partial_\mu\partial_\nu+m^2)\psi = 0
\end{eqnarray}
which must be the Klein-Gordon equation $(\partial^2 + m^2)\psi$ = 0
in our antisymmetric space-time (note it is the same as in flat space).
Therefore the Dirac algebra in this antisymmetric space 
is unchanged from the flat space case, $\ie$
 $\{\gamma^\mu,\gamma^\nu\}=2\eta^{\mu\nu}$ still holds with
the usual 4-dimensional matrices.

\subsection*{Pair Annihilation}
Starting from the matrix element in Eqn (\ref{amp}) and making the
substitutions $p_1\to p,~~ p_2 \to -p^\prime,~~ k_1 \to -k,~~ k_2 \to k^\prime$
gives
\begin{eqnarray}
iM&=&-ie^2\epsilon^*_\mu(k^\prime)\epsilon_\nu(k)\bar{u}(p^\prime)[\frac{\gamma^\mu(p\hspace{-0.2cm}\slash+k\hspace{-0.2cm}\slash+m)\gamma^\nu}{(p+k)^2-m^2}
+\frac{\gamma^\nu(p\hspace{-0.2cm}\slash-k\hspace{-0.2cm}\slash^\prime+m)\gamma^\mu}{(p-k^\prime)^2-m^2}]u(p)\nonumber\\
&=&-ie^2\epsilon_\mu^*(k^\prime)\epsilon_\nu(k)\bar{u}(p^\prime)[\frac{\gamma^\mu
k\hspace{-0.2cm}\slash\gamma^\nu+2\gamma^\mu\eta^{\nu\alpha}p_\alpha}{2p\cdot
k}+\frac{-\gamma^\nu
k\hspace{-0.2cm}\slash^\prime\gamma^\mu+2\gamma^\nu\eta^{\mu\alpha}p_\alpha}{-2p\cdot
k^\prime}]u(p)
\end{eqnarray}
From the kinematic definitions  of $k_\mu$ and $k^\prime_\mu$
we
can get $\epsilon^\mu(k)$ and $\epsilon^\mu(k^\prime)$, so that
\begin{eqnarray}
&&{1\over4}\sum_s|{\cal
M}|^2\nonumber\\
&&={e^4\over4}\sum_s\bigg\{g_{\mu\lambda}\epsilon^{*\lambda}(k^\prime)g_{\nu\varphi}\epsilon^\varphi(k)\bar{u}(p^\prime)\bigg[\frac{\gamma^\mu
k\hspace{-0.2cm}\slash\gamma^\nu+2\gamma^\mu\eta^{\nu\rho}p_\rho}{2p\cdot
k}+\frac{-\gamma^\nu
k\hspace{-0.2cm}\slash^\prime\gamma^\mu+2\gamma^\nu\eta^{\mu\rho}p_\rho}{-2p\cdot
k^\prime}\bigg]u(p)\bigg\}\nonumber\\
&&\times\bigg\{g_{\alpha\delta}\epsilon^{*\delta}(k^\prime)g_{\beta\theta}\epsilon^\theta(k)\bar{u}(p^\prime)\bigg[\frac{\gamma^\alpha
k\hspace{-0.2cm}\slash\gamma^\beta+2\gamma^\alpha\eta^{\beta\sigma}p_\sigma}{2p\cdot
k}+\frac{-\gamma^\beta
k\hspace{-0.2cm}\slash^\prime\gamma^\alpha+2\gamma^\beta\eta^{\alpha\sigma}p_\sigma}{-2p\cdot
k^\prime}\bigg]u(p)\bigg\}^\dag\nonumber\\
&&={e^4\over4}g_{\mu\lambda}g_{\nu\varphi}g_{\alpha\delta}g_{\beta\theta}\epsilon^{*\lambda}\epsilon^\delta(k^\prime)(k^\prime)\epsilon^{*\theta}(k)\epsilon^\varphi(k)\nonumber\\
&&\times
tr\bigg\{(p\hspace{-0.2cm}\slash^\prime+m)\bigg[\frac{\gamma^\mu
k\hspace{-0.2cm}\slash\gamma^\nu+2\gamma^\mu\eta^{\nu\rho}p_\rho}{2p\cdot
k}+\frac{-\gamma^\nu
k\hspace{-0.2cm}\slash^\prime\gamma^\mu+2\gamma^\nu\eta^{\mu\rho}p_\rho}{-2p\cdot
k^\prime}\bigg]\nonumber\\
&&\times(p\hspace{-0.2cm}\slash+m)\bigg[\frac{\gamma^\alpha
k\hspace{-0.2cm}\slash\gamma^\beta+2\gamma^\alpha\eta^{\beta\sigma}p_\sigma}{2p\cdot
k}+\frac{-\gamma^\beta
k\hspace{-0.2cm}\slash^\prime\gamma^\alpha+2\gamma^\beta\eta^{\alpha\sigma}p_\sigma}{-2p\cdot
k^\prime}\bigg]\bigg\}\nonumber\\
&&\equiv{e^4\over4}g_{\mu\lambda}g_{\nu\varphi}g_{\alpha\delta}g_{\beta\theta}\epsilon^{*\lambda}(k^\prime)\epsilon^\delta(k^\prime)\epsilon^{*\theta}(k)\epsilon^\varphi(k)\nonumber\\
&&\times\bigg[\frac{I}{(2p\cdot k)^2}+\frac{II}{(2p\cdot
k)(2p\cdot k^\prime)}+\frac{III}{(2p\cdot k)(2p\cdot
k^\prime)}+\frac{IV}{(2p\cdot k^\prime)^2}\bigg]
\end{eqnarray}
where
\begin{eqnarray}
I&=&tr\{(p\hspace{-0.2cm}\slash^\prime+m)(\gamma^\mu
k\hspace{-0.2cm}\slash\gamma^\nu+2\gamma^\mu\eta^{\nu\rho}p_\rho)(p\hspace{-0.2cm}\slash+m)(\gamma^\alpha
k\hspace{-0.2cm}\slash\gamma^\beta+2\gamma^\alpha\eta^{\beta\sigma}p_\sigma)\}\nonumber\\
&=&tr\{p\hspace{-0.2cm}\slash^\prime(\gamma^\mu
k\hspace{-0.2cm}\slash\gamma^\nu+2\gamma^\mu\eta^{\nu\rho}p_\rho)p\hspace{-0.2cm}\slash(\gamma^\alpha
k\hspace{-0.2cm}\slash\gamma^\beta+2\gamma^\alpha\eta^{\beta\sigma}p_\sigma)\}\nonumber\\
&=&tr\{p\hspace{-0.2cm}\slash^\prime\gamma^\mu
k\hspace{-0.2cm}\slash\gamma^\nu
p\hspace{-0.2cm}\slash\gamma^\alpha
k\hspace{-0.2cm}\slash\gamma^\beta\}\nonumber\\
&&+2tr\{p\hspace{-0.2cm}\slash^\prime\gamma^\mu
k\hspace{-0.2cm}\slash\gamma^\nu p\hspace{-0.2cm}\slash\gamma^\alpha\eta^{\beta\sigma}p_\sigma\}\nonumber\\
&&+2tr\{p\hspace{-0.2cm}\slash^\prime\gamma^\mu\eta^{\nu\rho}p_\rho
p\hspace{-0.2cm}\slash\gamma^\alpha
k\hspace{-0.2cm}\slash\gamma^\beta\}\nonumber\\
&&+4tr\{p\hspace{-0.2cm}\slash^\prime\gamma^\mu\eta^{\nu\rho}p_\rho
p\hspace{-0.2cm}\slash \gamma^\alpha\eta^{\beta\sigma}p_\sigma\}\\
IV&=&I(k\rightarrow k^\prime)\\
II&=&tr\{(p\hspace{-0.2cm}\slash^\prime+m)(\gamma^\mu
k\hspace{-0.2cm}\slash\gamma^\nu+2\gamma^\mu\eta^{\nu\rho}p_\rho)(p\hspace{-0.2cm}\slash+m)(-\gamma^\beta
k\hspace{-0.2cm}\slash^\prime\gamma^\alpha+2\gamma^\beta\eta^{\alpha\sigma}p_\sigma)\}\nonumber\\
&=&-tr\{p\hspace{-0.2cm}\slash^\prime\gamma^\mu
k\hspace{-0.2cm}\slash\gamma^\nu
p\hspace{-0.2cm}\slash\gamma^\beta
k\hspace{-0.2cm}\slash^\prime\gamma^\alpha\}\nonumber\\
&&+2tr\{p\hspace{-0.2cm}\slash^\prime\gamma^\mu
k\hspace{-0.2cm}\slash\gamma^\nu p\hspace{-0.2cm}\slash\gamma^\beta\eta^{\alpha\sigma}p_\sigma\}\nonumber\\
&&-2tr\{p\hspace{-0.2cm}\slash^\prime\gamma^\mu\eta^{\nu\rho}p_\rho
p\hspace{-0.2cm}\slash\gamma^\beta
k\hspace{-0.2cm}\slash^\prime\gamma^\alpha\}\nonumber\\
&&+4tr\{p\hspace{-0.2cm}\slash^\prime\gamma^\mu\eta^{\nu\rho}p_\rho
p\hspace{-0.2cm}\slash\gamma^\beta\eta^{\alpha\sigma}p_\sigma\}\\
III&=&II
\end{eqnarray}
After some calculation, we get
\begin{eqnarray}
I&=&32[(p^\prime\star\epsilon)(p\star\epsilon)(k\star\epsilon)(p\star\epsilon)-(p^\prime\star\epsilon)(p\cdot
k)(\epsilon\star\epsilon)(p\star\epsilon)-(p^\prime\cdot
k)(\epsilon\star\epsilon)(p\star\epsilon)(p\star\epsilon)\nonumber\\
&&+(p^\prime\cdot
k)(p\star\epsilon)(k\star\epsilon)(p\star\epsilon)+(p^\prime\star\epsilon)(k\star\epsilon)(p\star\epsilon)(p\star\epsilon)-(p^\prime\cdot
p)(k\star\epsilon)(\epsilon\star\epsilon)(p\star\epsilon)\nonumber\\
&&+(p^\prime\star\epsilon)(p\star\epsilon)(p\star\epsilon)(p\star\epsilon)-(p^\prime\cdot
p)(p\star\epsilon)(p\star\epsilon)(\epsilon\star\epsilon)+(p^\prime\star\epsilon)(k\star\epsilon)(p\star\epsilon)(k\star\epsilon)\nonumber\\
&&-(p^\prime\cdot
p)(k\star\epsilon)(k\star\epsilon)(\epsilon\star\epsilon)]\nonumber\\
&=&32[(p^\prime\star\epsilon)(p\star\epsilon)^2(k\star\epsilon)-(p^\prime\star\epsilon)(p\cdot
k)(\epsilon\star\epsilon)(p\star\epsilon)-(p^\prime\cdot
k)(\epsilon\star\epsilon)(p\star\epsilon)^2\nonumber\\
&&+(p^\prime\cdot
k)(p\star\epsilon)^2(k\star\epsilon)+(p^\prime\star\epsilon)(k\star\epsilon)(p\star\epsilon)^2-(p^\prime\cdot
p)(k\star\epsilon)(\epsilon\star\epsilon)(p\star\epsilon)\nonumber\\
&&+(p^\prime\star\epsilon)(p\star\epsilon)^3-(p^\prime\cdot
p)(p\star\epsilon)^2(\epsilon\star\epsilon)+(p^\prime\star\epsilon)(k\star\epsilon)^2(p\star\epsilon)\nonumber\\
&&-(p^\prime\cdot
p)(k\star\epsilon)^2(\epsilon\star\epsilon)]\\
IV&=&32[(p^\prime\star\epsilon)(p\star\epsilon)^2(k\star\epsilon)-(p^\prime\star\epsilon)(p\cdot
k^\prime)(\epsilon\star\epsilon)(p\star\epsilon)-(p^\prime\cdot
k^\prime)(\epsilon\star\epsilon)(p\star\epsilon)^2\nonumber\\
&&+(p^\prime\cdot
k^\prime)(p\star\epsilon)^2(k^\prime\star\epsilon)+(p^\prime\star\epsilon)(k^\prime\star\epsilon)(p\star\epsilon)^2-(p^\prime\cdot
p)(k^\prime\star\epsilon)(\epsilon\star\epsilon)(p\star\epsilon)\nonumber\\
&&+(p^\prime\star\epsilon)(p\star\epsilon)^3-(p^\prime\cdot
p)(p\star\epsilon)^2(\epsilon\star\epsilon)+(p^\prime\star\epsilon)(k^\prime\star\epsilon)^2(p\star\epsilon)\nonumber\\
&&-(p^\prime\cdot
p)(k^\prime\star\epsilon)^2(\epsilon\star\epsilon)]\\
II&=&16[(p^\prime\star\epsilon)(p\star\epsilon)^2(k\star\epsilon)-(p^\prime\star\epsilon)(p\cdot
k)(\epsilon\star\epsilon)(p\star\epsilon)-(p^\prime\cdot
k)(\epsilon\star\epsilon)(p\star\epsilon)^2\nonumber\\
&&+(p^\prime\cdot
k)(p\star\epsilon)^2(k\star\epsilon)+(p^\prime\star\epsilon)(k\star\epsilon)(p\star\epsilon)^2-(p^\prime\cdot
p)(k\star\epsilon)(\epsilon\star\epsilon)(p\star\epsilon)\nonumber\\
&&+(p^\prime\star\epsilon)(p\star\epsilon)^3-(p^\prime\cdot
p)(p\star\epsilon)^2(\epsilon\star\epsilon)+(p^\prime\star\epsilon)(k\star\epsilon)(k^\prime\star\epsilon)(p\star\epsilon)\nonumber\\
&&-(p^\prime\cdot
p)(k\star\epsilon)(k^\prime\star\epsilon)(\epsilon\star\epsilon)]\\
III&=&II
\end{eqnarray}
in the above, we take the polarization of the photons to be real 
and used the definition
\begin{eqnarray}
&&k\cdot p=k_\mu\eta^{\mu\nu}p_\nu\\
&&p\star\epsilon=p_\mu\eta^{\mu\nu}g_{\nu\alpha}\epsilon^\alpha\\
&&\epsilon\star\epsilon=g_{\mu\alpha}\epsilon^\alpha\eta^{\mu\nu}g_{\nu\beta}\epsilon^\beta
\end{eqnarray}

Now define
\begin{eqnarray}
&&p_\mu=(E,0,0,E)~~~~~~~~p_\mu^\prime=(E,0,0,-E)\\
&&k_\mu=(E,E\sin\theta\cos\varphi,E\sin\theta\sin\varphi,E\cos\theta)\\
&&k^\prime_\mu=(E,-E\sin\theta\cos\varphi,-E\sin\theta\sin\varphi,-E\cos\theta)
\end{eqnarray}
then, we get
\begin{eqnarray}
&&\epsilon^{1\mu}=(0,\cos\varphi\cos\theta,\sin\varphi\cos\theta,-\sin\theta)\\
&&\epsilon^{2\mu}=(0,-\sin\varphi,\cos\varphi,0)
\end{eqnarray}

Now rewrite the metric matrix as
\begin{eqnarray}
(a_{\mu\nu})=\left(%
\begin{array}{cccc}
  1 & a & b & c \\
  -a & -1 & d & h \\
  -b & -d & -1 & r \\
  -c & -h & -r & -1 \\
\end{array}%
\right)
\end{eqnarray}
where $a,b,c,d,h,r$ are all much less than unity.
Note that in the case $a=b=c=d=h=r$ we would have
\begin{eqnarray}
p\cdot k&=&p^\prime\cdot k^\prime=E^2(1-\cos\theta)\\
p\cdot k^\prime&=&p^\prime\cdot k=E^2(1+\cos\theta)\\
p\star\epsilon^1&=&-E(1+a)\sin\theta+2aE\cos\theta(\sin\varphi+\cos\varphi)\\
p\star\epsilon^2&=&2aE(\sin\varphi+\cos\varphi)\\
p^\prime\star\epsilon^1&=&E(1-a)\sin\theta\\
p^\prime\star\epsilon^2&=&0\\
k\star\epsilon^1&=&E[a(1+\cos\theta)(\sin\varphi+\cos\varphi)+(1-a)\sin\theta\cos\theta\sin\varphi\cos\varphi\nonumber\\
&&-\sin\theta\cos\theta(1-\cos\varphi)-a\sin\theta+\sin\theta\cos\theta\sin^2\varphi]\\
k\star\epsilon^2&=&E[a(1+\cos\theta)(\sin\varphi+\cos\varphi)+a\sin\theta(\sin^2\varphi-\cos^2\varphi)+\sin\theta\sin^2\varphi]\\
k^\prime\star\epsilon^1&=&E[-a(1-\cos\theta)(\sin\varphi+\cos\varphi)-a\sin\theta]\\
k^\prime\star\epsilon^2&=&E[a(1-\cos\theta)(\sin\varphi+\cos\varphi)+a\sin\theta\cos2\varphi-\sin\theta\sin2\varphi]\\
\epsilon^1\star\epsilon^1&=&-2a^2\sin2\theta\cos\varphi-(1+a^2)[\sin^2\theta+\cos^2\theta(\sin\varphi+\cos\varphi)]\\
\epsilon^2\star\epsilon^2&=&-(1+a^2)
\end{eqnarray}
 To a first
approximation we need only keep terms in the scattering cross section 
proportional to any {\it one} of $a^2, b^2, c^2, d^2, h^2, r^2$.
This gives 
\subsection*{\bf $a$-dependent terms}
\begin{eqnarray}
I&=&32E^4\bigg\{(\cos\theta\cos\varphi+\sin\theta)(a\cos\theta\cos\varphi-\sin\theta)\bigg[(a\cos\theta\cos\varphi+a\cos\theta\cos\varphi+\sin\theta)^2\nonumber\\
&&+(-\cos\theta+\cos\varphi\sin\theta+1)\bigg]+2\bigg[(a\cos\theta\cos\varphi)^2+(a\cos\theta\cos\varphi+\sin\theta)(a\cos\theta\cos\varphi)\nonumber\\
&&+(a\cos\theta\cos\varphi+\sin\theta)^2\bigg]\bigg\}
\end{eqnarray}
\begin{eqnarray}
II&=&16E^4\bigg\{(a\cos\theta\cos\varphi+\sin\theta)\bigg[(a\cos\theta\cos\varphi-\sin\theta)\{(a\cos\theta\cos\varphi)^2\nonumber\\
&&+3(a\cos\theta\cos\varphi+\sin\theta)(a\cos\theta\cos\varphi)+(a\cos\theta\cos\varphi+\sin\theta)^2\nonumber\\
&&+(-\cos\theta+a\cos\varphi\sin\theta+1)\}+(a\cos\theta\cos\varphi+1)(a\cos\theta\cos\varphi+\sin\theta)\nonumber\\
&&\times(\cos\theta+a\cos\varphi\sin\theta+1)\bigg]+2\bigg[(a\cos\theta\cos\varphi+\sin\theta)^2\nonumber\\
&&+(a\cos\theta\cos\varphi)\{a\cos\theta\cos\varphi+(a\cos\theta\cos\varphi+\sin\theta)\}\bigg]\bigg\}
\end{eqnarray}
\begin{eqnarray}
IV&=&32E^4\bigg\{(a\cos\theta\cos\varphi+\sin\theta)(a\cos\theta\cos\varphi-\sin\theta)\{a\cos\theta\cos\varphi+(a\cos\theta\cos\varphi+\sin\theta)\}^2\nonumber\\
&&+(\cos\theta-a\cos\varphi\sin\theta+1)-(a\cos\theta\cos\varphi+1)(a\cos\theta\cos\varphi+\sin\theta)\nonumber\\
&&\times(\cos\theta+a\cos\varphi\sin\theta-1)+2\{(a\cos\theta\cos\varphi)^2+(a\cos\theta\cos\varphi+\sin\theta)\nonumber\\
&&\times(a\cos\theta\cos\varphi)+(a\cos\theta\cos\varphi+\sin\theta)^2\}\bigg\}
\end{eqnarray}

\subsection*{\bf $b$-dependent terms}
\begin{eqnarray}
I&=&32E^4\bigg\{(\sin\theta+b\cos\theta\sin\varphi)(b\cos\theta\sin\varphi-e\sin\theta)\bigg[\{b\cos\theta\sin\varphi+(\sin\theta+b\cos\theta\sin\varphi)\}^2\nonumber\\
&&+(-\cos\theta+b\sin\theta\sin\varphi+1)\bigg]+(b\cos\theta\sin\varphi+1)(\sin\theta+b\cos\theta\sin\varphi)\nonumber\\
&&\times(\cos\theta+b\sin\theta\sin\varphi+1)+2\bigg[(b\cos\theta\sin\varphi)^2+(\sin\theta+b\cos\theta\sin\varphi)(b\cos\theta\sin\varphi)\bigg]\nonumber\\
&&+(\sin\theta+b\cos\theta\sin\varphi)^2\bigg\}\nonumber\\
II&=&16E^4\bigg\{(\sin\theta+b\cos\theta\sin\varphi)\bigg[(b\cos\theta-e\sin\theta)\{(b\cos\theta\sin\varphi)^2\nonumber\\
&&+3(\sin\theta+b\cos\theta\sin\varphi)(b\cos\theta\sin\varphi)+(\sin\theta+b\cos\theta\sin\varphi)^2\nonumber\\
&&+(-\cos\theta+b\sin\theta\sin\varphi+1)\}+(b\cos\theta\sin\varphi+1)(\sin\theta+b\cos\theta\sin\varphi)\nonumber\\
&&\times(\cos\theta+b\sin\theta\sin\varphi+1)\bigg]+2\bigg[(\sin\theta+b\cos\theta\sin\varphi)^2\nonumber\\
&&+(b\cos\theta\sin\varphi)\{b\cos\theta\sin\varphi+(\sin\theta+b\cos\theta\sin\varphi)\}\bigg]\bigg\}\nonumber\\
IV&=&32E^4\bigg\{(\sin\theta+b\cos\theta\sin\varphi)\bigg[(b\cos\theta\sin\varphi-\sin\theta)\{[b\cos\theta\sin\varphi+(\sin\theta+b\cos\theta\sin\varphi)]^2\nonumber\\
&&+(\cos\theta-b\sin\theta\sin\varphi+1)\}-(b\cos\theta\sin\varphi+1)(\sin\theta+b\cos\theta\sin\varphi)\nonumber\\
&&\times(\cos\theta+b\sin\theta\sin\varphi-1)\bigg]+2\{(b\cos\theta\sin\varphi)^2+(\sin\theta+b\cos\theta\sin\varphi)\nonumber\\
&&\times(b\cos\theta\sin\varphi)+(\sin\theta+b\cos\theta\sin\varphi)^2\}\bigg\}
\end{eqnarray}

\subsection*{\bf $c$-dependent terms}
\begin{eqnarray}
I&=&32E^4\bigg\{(c-1)\sin\theta\bigg[(c+1)(\cos\theta+1)(c-1)\sin\theta(-c\sin\theta+1)\nonumber\\
&&+\{[(c-1)\sin\theta+c\sin\theta]^2-(c-1)(\cos\theta-1)\}(c+1)\sin\theta\bigg]\nonumber\\
&&-2(c-1)\{[(c-1)\sin\theta]^2+c\sin\theta(c-1)\sin\theta+(c\sin\theta)^2\}\bigg\}\nonumber\\
II&=&E^4\bigg\{-(c-1)\sin\theta\bigg[-(c+1)(\cos\theta+1)(c-1)\sin\theta(-c\sin\theta+1)\nonumber\\
&&-\{[(c-1)\sin\theta]^2+3c\sin\theta(c-1)\sin\theta+(c\sin\theta)^2\nonumber\\
&&+(c-1)(\cos\theta-1)\}(c+1)\sin\theta\bigg]-2(c-1)\{[(c-1)\sin\theta]^2\nonumber\\
&&+c\sin\theta[(c-1)\sin\theta+c\sin\theta]\}\bigg\}\nonumber\\
IV&=&32E^4\bigg\{-(c-1)\sin\theta\bigg[(2(c+1)\sin^2({\theta\over2}))(-(c-1)\sin\theta)(-c\sin\theta+1)\nonumber\\
&&-\{[(c-1)\sin\theta+c\sin\theta]^2-2(c-1)\cos^2({\theta\over2})\}(c+1)\sin\theta\bigg]\nonumber\\
&&-2(c-1)\{[(c-1)\sin\theta]^2+c\sin\theta(c-1)\sin\theta+(c\sin\theta)^2\}\bigg\}
\end{eqnarray}
\subsection*{\bf $d$-dependent terms}

These all cancel.\\

\subsection*{\bf $h$-dependent terms}
\begin{eqnarray}
I&=&32E^4\bigg\{(\sin\theta-h\cos\theta\cos\varphi)\bigg[(h\cos\theta\cos\varphi-\sin\theta)\{[-h\cos\varphi+(\sin\theta-h\cos\theta\cos\varphi)]^2\nonumber\\
&&-(\cos\theta+h\cos\varphi\sin\theta-1)\}+(-h\cos\varphi+1)(\sin\theta-h\cos\theta\cos\varphi)\nonumber\\
&&\times(\cos\theta+h\cos\varphi\sin\theta+1)\bigg]+2\{(-h\cos\varphi)^2\nonumber\\
&&+(\sin\theta-h\cos\theta\cos\varphi)(-h\cos\varphi)+(\sin\theta-h\cos\theta\cos\varphi)^2\}\bigg\}\nonumber\\
II&=&16E^4\bigg\{(\sin\theta-h\cos\theta\cos\varphi)\bigg[(h\cos\theta\cos\varphi-\sin\theta)\{(\sin\theta-h\cos\theta\cos\varphi)^2\nonumber\\
&&+(-h\cos\varphi)[h\cos\varphi+3(\sin\theta-h\cos\theta\cos\varphi)]-(\cos\theta+h\cos\varphi\sin\theta-1)\}\nonumber\\
&&+(-h\cos\varphi+1)(\sin\theta-h\cos\theta\cos\varphi)(\cos\theta+h\cos\varphi\sin\theta+1)\bigg]\nonumber\\
&&+2\{(\sin\theta-h\cos\theta\cos\varphi)^2-h\cos\varphi(h\cos\varphi-h\cos\theta\cos\varphi)\}\bigg\}\nonumber\\
IV&=&32E^4\bigg\{(\sin\theta-h\cos\theta\cos\varphi)\bigg[-(h\cos\varphi+1)(\sin\theta-h\cos\theta\cos\varphi)\nonumber\\
&&\times(\cos\theta+h\cos\varphi\sin\theta-1)+(h\cos\theta\cos\varphi-\sin\theta)\nonumber\\
&&\times\{[h\cos\varphi+(\sin\theta-h\cos\theta\cos\varphi)]^2+(\cos\theta+h\sin\theta\cos\varphi+1)\}\bigg]\nonumber\\
&&+2\{(h\cos\varphi)^2+(\sin\theta-h\cos\theta\cos\varphi)(h\cos\varphi)+(\sin\theta-h\cos\theta\cos\varphi)^2\}\bigg\}
\end{eqnarray}

\subsection*{\bf $r$-dependent terms}
\begin{eqnarray}
I&=&32E^4\bigg\{(\sin\theta-r\cos\theta\sin\varphi)\bigg[(r\cos\theta\sin\varphi-\sin\theta)\{(-r\sin\varphi+(\sin\theta-r\cos\theta\sin\varphi))^2\nonumber\\
&&-(\cos\theta+r\sin\theta\sin\varphi-1)\}+(-r\sin\varphi+1)(\sin\theta-r\cos\theta\sin\varphi)\nonumber\\
&&\times(\cos\theta+r\sin\theta\sin\varphi+1)\bigg]+2\{(-r\sin\varphi)^2+(\sin\theta-r\cos\theta\sin\varphi)(-r\sin\varphi)\nonumber\\
&&+(\sin\theta-r\cos\theta\sin\varphi)^2\}\bigg\}\nonumber\\
II&=&16E^4\bigg\{(\sin\theta-r\cos\theta\sin\varphi)\bigg[(r\cos\theta\sin\varphi-\sin\theta)\{(\sin\theta-r\cos\theta\sin\varphi)^2\nonumber\\
&&-r\sin\varphi[r\sin\varphi+3(\sin\theta-r\cos\theta\sin\varphi)]-(\cos\theta+r\sin\theta\sin\varphi-1)\}\nonumber\\
&&+(-r\sin\varphi+1)(\sin\theta-r\cos\theta\sin\varphi)(\cos\theta+r\sin\theta\sin\varphi+1)\bigg]\nonumber\\
&&+2\{(\sin\theta-r\cos\theta\sin\varphi)^2+(-r\sin\varphi)[r\sin\varphi+(\sin\theta-r\cos\theta\sin\varphi)]\}\bigg\}\nonumber\\
IV&=&32E^4\bigg\{(\sin\theta-r\cos\theta\sin\varphi)\bigg[-(r\sin\varphi+1)(\sin\theta-r\cos\theta\sin\varphi)\nonumber\\
&&\times(\cos\theta+r\sin\theta\sin\varphi-1)+(r\cos\theta\sin\varphi-\sin\theta)\{[r\sin\varphi+(\sin\theta-r\cos\theta\sin\varphi)]^2\nonumber\\
&&+(\cos\theta+r\sin\theta\sin\varphi+1)\}\bigg]+2\{(r\sin\varphi)^2+(\sin\theta-r\cos\theta\sin\varphi)(r\sin\varphi)\nonumber\\
&&+(\sin\theta-r\cos\theta\sin\varphi)^2\}\bigg\}\nonumber\\
\end{eqnarray}

where we have removed terms dependent on two or more of
$a,b,c,d,h,r$.
Now, simplifying by setting $a^2= b^2= c^2= d^2= h^2= r^2$
and after some algebra, 
\begin{eqnarray}
I&=&-32{aE^4\over8}\{\sin\theta(1+8\cos2\theta)(\sin\varphi+\cos\varphi)\}\nonumber\\
&&-32{a^2E^4\over4}\{2\cos\theta+\cos(\theta-4\varphi)\}\\
IV&=&32a\{\sin\theta[2(\sin\varphi+\cos\varphi)\cos^3\theta-{1\over2}(\sin\varphi+\cos\varphi)]\cos[2(\theta-\varphi)]\}\nonumber\\
&&+32a^2\{2\sin\theta(1-\cos\varphi)\cos^2\varphi(\sin\varphi+\cos\varphi)\}\\
II&=&16a\{\sin\theta[2(\sin\varphi+\cos\varphi-1)(\sin{3\varphi\over2}+\cos{\varphi\over2})^2\cos^3\theta+\sin^2\theta\sin\varphi]\}\nonumber\\
&&-16a^2\{\cos\theta\sin^2\varphi[\cos\theta(\sin\varphi+\cos\varphi)-\sin^2\theta]\}\\
III&=&II
\end{eqnarray}
At last, we have
\begin{eqnarray}
&&{1\over4}\sum_s|{\cal
M}|^2\nonumber\\
&&=\mbox{ordinary theory}\nonumber\\
&&+8a{e^4\over4\sin^4\theta}\bigg\{-{1\over8}(1+\cos\theta)^2\{\sin\theta(1+8\cos2\theta)(\sin\varphi+\cos\varphi)\}\nonumber\\
&&+\sin^3\theta[2(\sin\varphi+\cos\varphi-1)(\sin{3\varphi\over2}+\cos{\varphi\over2})^2\cos^3\theta+\sin^2\theta\sin\varphi]\nonumber\\
&&+(1-\cos\theta)^2\sin\theta[2(\sin\varphi+\cos\varphi)\cos^3\theta-{1\over2}(\sin\varphi+\cos\varphi)]\cos[2(\theta-\varphi)]\bigg\}\nonumber\\
&&+8a^2{e^4\over4\sin^4\theta}\bigg\{-{1\over4}(1+\cos\theta)^2\sin\theta(1+8\cos2\theta)(\sin\varphi+\cos\varphi)\nonumber\\
&&-\sin^2\theta\{\cos\theta\sin^2\varphi[\cos\theta(\sin\varphi+\cos\varphi)-\sin^2\theta]\}\nonumber\\
&&+(1-\cos\theta)^2\{2\sin\theta(1-\cos\varphi)\cos^2\varphi(\sin\varphi+\cos\varphi)\}\bigg\}\nonumber\\
&&=\mbox{ordinary theory}\nonumber\\
&&+8a{e^4\over4\sin^3\theta}\bigg\{-{1\over8}(1+\cos\theta)^2(1+8\cos2\theta)(\sin\varphi+\cos\varphi)\nonumber\\
&&+\sin^2\theta[2(\sin\varphi+\cos\varphi-1)(\sin{3\varphi\over2}+\cos{\varphi\over2})^2\cos^3\theta+\sin^2\theta\sin\varphi]\nonumber\\
&&+(1-\cos\theta)^2[2(\sin\varphi+\cos\varphi)\cos^3\theta-{1\over2}(\sin\varphi+\cos\varphi)]\cos[2(\theta-\varphi)]\bigg\}\nonumber\\
&&+8a^2{e^4\over4\sin^3\theta}\bigg\{-{1\over4}(1+\cos\theta)^2(1+8\cos2\theta)(\sin\varphi+\cos\varphi)\nonumber\\
&&-\sin\theta\{\cos\theta\sin^2\varphi[\cos\theta(\sin\varphi+\cos\varphi)-\sin^2\theta]\}\nonumber\\
&&+(1-\cos\theta)^2\{2(1-\cos\varphi)\cos^2\varphi(\sin\varphi+\cos\varphi)\}\bigg\}
\end{eqnarray}
\begin{eqnarray}
\end{eqnarray}
\begin{eqnarray}
\end{eqnarray}

\bibliography{all.bib}
\bibliographystyle{unsrt}

\end{document}